# SECURE TRANSMISSION IN WIRELESS SENSOR NETWORKS DATA USING LINEAR KOLMOGOROV WATERMARKING TECHNIQUE

Bambang Harjito[1] and Vidyasagar Potdar[2]

[1]Department of informatics, Mathematics and Natural Science Faculty, Sebelas Maret University, Surakarta, Indonesia
`bambang_harjito@staff.uns.ac.id`
[2]School of Information System, Curtin University, Perth, Australia
`v.potdar@curtin.edu.au`

## ABSTRACT

*In Wireless sensor networks (WSNs), All communications between different nodes are sent out in a broadcast fashion. These networks are used in a variety of applications including military, environmental, and smart spaces. Sensors are susceptible to various types of attack, such as data modification, data insertion and deletion, or even physical capture and sensor replacement. Hence security becomes important issue in WSNs. However given the fact that sensors are resources constrained, hence the traditional intensive security algorithms are not well suited for WSNs. This makes traditional security techniques, based on data encryption, not very suitable for WSNs. This paper proposes Linear Kolmogorov watermarking technique for secure data communication in WSNs. We provide a security analysis to show the robustness of the proposed techniques against various types of attacks. This technique is robust against data deletion, packet replication and Sybil attacks*

## 1. INTRODUCTION

Wireless Sensor Networks (WSNs) have the capability for sensing, processing and wireless communication all built into a tiny embedded device [1]. This type of network has drawn increasing interest in the research community over the last few years. This is driven by theoretical and practical problems in embedded operating systems, network protocols, wireless communications and distributed signal processing. The primary function of WSNs is to collect and disseminate critical data that characterize the physical phenomena within the target area.

We know that WSN nodes have low power supply and limited computational capability because they operate on batteries. Given their limited power supply it becomes challenges to use store for ensuring security. There are numerous security dimensions like authenticity, integrity, copyright data protection. Watermarking techniques are been investigated for addressing some of these issues like tampering, data authentication, copyright and detection etc. Watermarking algorithms are shown to be less energy demanding and the recent literature shows that incorporate watermarking in WSN is feasible. Hence the research in the area of watermarking and WSN is becoming increasingly important. Watermarking technique is a lightweight technique that was used traditionally for providing copyright protection for multimedia data like images and video clips. Watermarking algorithms are much lighter and require less battery power and processing capabilities than cryptographic-based algorithms. Another advantage for the watermarking-based algorithms is that the watermark is embedded directly into the sensor data; there is no increase in the payload. While cryptography provides no protection after the content is decrypted, watermarking provides protection in secrecy at all times because the watermark is an inseparable constituent part of the host media [6-8]. Hence the research in the area of watermarking and WSN is becoming increasingly important. With the concept of cyber physical system, i.e., on web of things this research is becoming main stream and the importance of this research has become even more significant. The objective of this paper we present on secure data trnasmittion in WSNs using watermarking technique.

## 2. RELATED WORKS

In the last few years, there are many researches who studies on digital watermarking technique for normal data types for example texts, images, audios, videos. and even relational databases [2-4] But there are only a few research works on digital watermarking techniques for WNSs [5] [6, 7]. Feng, J.P et.al [5] developed the first system of watermarking technique to embed crypto logically encoded authorship signatures into data and information acquired by wireless embedded sensor networks. Sion et.al [6] provided copyright protection to data stream owners and authorized users. Consider the case where a stream is generated and safely transmitted from the sensors to the base station. A watermark is applied to the stream at the base station. The data are then transmitted to an authorized user. The owner and authorized users need a way to show that the data were generated by them and they want to prove that the stream was illegally obtained by the attacker. One commonly accepted way to prove ownership is the use of embedded watermarks. This technique works by embedding a watermark bit into major extremes, which are extremes that will survive any uniform sampling. F. Koushanfar et.al [8] present an active watermarking technique that can be used on the data that is processing during the common sensor fusion application from sensor of different modalities. Xiaoet.al [9]proposed a watermarking technique for protecting copyright by taking advantage of the characteristic of the sending time. Based on digital watermarking, Zhang, W, et.al [10] proposed an end-to-end, watermark statistical approach for data authentication that provides inherent support for in-network processing. In this technique authentication information is modulated as watermark and then is embedded to the sensory data at the sensor nodes. Communication protocol for WSN is introduced by Xuejun R et.al [11] for sensitive data transmission. The technique use sensitive information as watermark. The watermark is then embedded into sensory data in the sensor nodes. A threshold is used for avoiding the alteration of the lowest to make a big influence to sensory data's precision. Kamel et.al [12] introduced a technique for providing data integrity. This technique based on distortion free watermarking embeds the watermark in the order the data element so that it does not cause distortion to the data.

Usually there are two main purposes for watermarking. One of is to protect the copyright of the author. The other is to provide data integrity and to do authentication by using user's identity as watermark information. Compared with authentication schemes based on public key ciphers, the watermarking based authentication has the advantages of lower computational complexity and being invisible to adversaries. In fact, besides these purposes, watermarking technique can also be used to transmit some secret information through unsecure channels without encryption. The table 1 shows their approach and their purpose many researchers who work on the watermarking technique for WSNs.

Although some research works attempted to apply digital watermarking technique into wireless sensor networks for copyright protection, authentication and integrity purposes, most of existing studies were only limited to secure data communication. No watermarking based secure data communication method has been found in related works. Therefore the purpose of this paper is that it presents secure data transmission in WSNs using watermarking technique

Table 1 Watermark embedding approaches and their purpose

| Author | Watermark embedding technique | Purpose |
|---|---|---|
| Feng et al. [5] | Adding watermark constraint to processing step during network operation | Copyright protection |
| Sion et al. [6] | Selection criteria using MSB | Copyright protection |
| Koushanfar et al. [8] | Adding watermark constraint to processing step during network operation | Copyright protection |
| Xiao et al. [9] | By modification the embedding bit of each packet. LSB | Copyright protection |
| Zhang et al. [10] | The watermark sensory data, $d(x,y) = w(x,y)+o(x,y)$, $w(x,y)$ is the watermark for sensor node and $O(x,y)$ is sensory data | Authentication |
| Xuejun et al. [11] | IIS = input integer stream, IBs=input binary stream. T = Threshold, If IIS ≥T<br>"IBS=1" become "IBS=0"<br>Else "IBS=0" become "IBS=0" | Authentication |
| Kamel et al. [13] | Concatenation of the current group hash value group $g_i$ and next group hash value group $g_{i+1}$. $W_i$ = HASH (K // $g_i$ // SN) SN = serial number | Integrity |

## 3. AN OVERVIEW OF DIGITAL WATERMARK

Watermarking technique is the process of embedding information which allows an individual to add hidden copyright notices or other verification messages to digital audio, video, or image signals and documents object [14-16]. Such hidden message is a group of bits describing information pertaining to the signal or to the author of the signal. The signal may be audio, pictures or video, for example, if the signal is copied, then the information is also carried in the copy. Watermarking seeks to embed a unique piece of data into the cover medium. The specific requirements of each watermarking technique may vary with the applications and there is no universal watermarking technique that satisfies all the requirements completely for all application.

Watermarking system as a communication task consists of three main stages: watermark generation process, watermark embedding process that including information transmission and possible attacks through the communication channel and detecting process that watermark retrieval

### 3.1. Watermark generation process

Generation process is the first step and a very critical of the process. The requirements of watermark generation process are unique and complexity. The watermark message contains information that must be unique such as simple text [5] [8] The key embedding is also unique in order to make a secrecy key such as binary stream [13] [17] [18] [9] [19] and pseudorandom sequence [10]. Both the watermark message and the key embedding are as input and they then are processed in the watermark generator to produce a watermark signal. Examples of the watermark generator are Hash function [13] [5] [8] [17] [18] [19] and product function . The watermark signal is a kind of signal or pattern that can be embedded into cover medium. There two types of watermark signal, i.e., meaningful and meaningless watermark. Examples of the watermark meaningful are logo

### 3.2. Watermark embedding Process

Embedding process is the second step of the watermarking system. This process is undertaken by an embedder and can be done in the transform domain such as Discrete Cosine Transform

(DCT), Discrete Fourier Transform (DFT), Fast Fourier Transform (FFT) and Discrete Wavelet Transform (DWT). The embedder combines the cover medium, the watermark signal, the sensed data and key embedding and it then creates watermarked cover medium. Examples of the cover medium are packed data, text, image, audio signal and video. The watermarked cover medium is perceptually identical to the cover medium. The watermarked cover medium is then transmitted by the sender through the unsecure communication channel such as wireless and radio channel. During transmission, there is anything that interfere in the communication process such noise, decreasing the quality of transmitting and a watermarked cover medium dropped. The other thing is that watermark attacks such as cropping, compression, and filtering, the aim of this attack is removed the watermark signal from the watermarked cover medium

### 3.3. Detecting and Extracting Process

The end of the watermarking system detects or extracts process that is a crucial part because the sender can identify and provide information to the intended receiver. The detecting or extracting is undertaken by a detector. The detecting process consists of an extraction unit to first extract the watermark signal and later compare it with the cover medium or not inserted. The extracting process can be divided into two phases, locating the watermark and recovering the watermark information. There are two types detection: Informed detection and blind detection according whether the cover medium is needed or not in the detection process. For informed detection which means the cover medium such as a packet data, original image and original signal, the watermarking system is called private watermarking. For the blind detection that does not need the cover medium is used for detection, the watermarking system is called a public watermarking.

## 4. PROBLEM DESCRIPTION

In application the wireless sensor networks, all communication between different nodes are send in broadcast fashion through communication channel where any node become attack target with external and internal security risk including eavesdropping, leak, temper, disrupt and other. In the special application fields, if the data transmission is not reliable, the security of the whole networks will be affected. Secure data transmission between sensors nodes have become important issue because an attacker can easily eavesdrop on, inject and manipulate a sensor node. Secure data sensor networks use many cryptographic algorithms. These techniques need thousands or even millions of multiplication instructions in order to perform operations [20-24]. The essence of the public key encryption for WSNs is keeping information the plain packet data secret namely securing communication in the presence of attackers, verifying authenticity of trusted parties and maintaining transaction integrity. In the previous section, we conducted an in-depth literature survey of watermarking approach in WSNs solution and their purposes and we identified that only limited presented. a solution to the address of issues. This gives us the rationale to present our solution secure data transmission model based on watermarking technique

## 5. PROPOSED MODEL WATERMARKING TECHNIQUE

In this section, we give a general overview of our solution watermarking technique to protect the reliability of data transmissions. The secure data communication model based on watermarking is illustrated in Figure 1. According to the model, this model consists of four steps : (a) cover medium process (b) Watermark generator process (c) Embedder process (d) Detecting or extracting process. The cover medium process is the process to generate a cover medium by using an atomic trilateration process. The watermark generator process is to create watermark constraint and message sensed data. This process requires a sensed data whereas the data through the LFSR process, partitioned process and Kolmogorof rule process. The embedder process is the process to generate a cover medium watermarked and the process of detecting is to detect the watermark signal

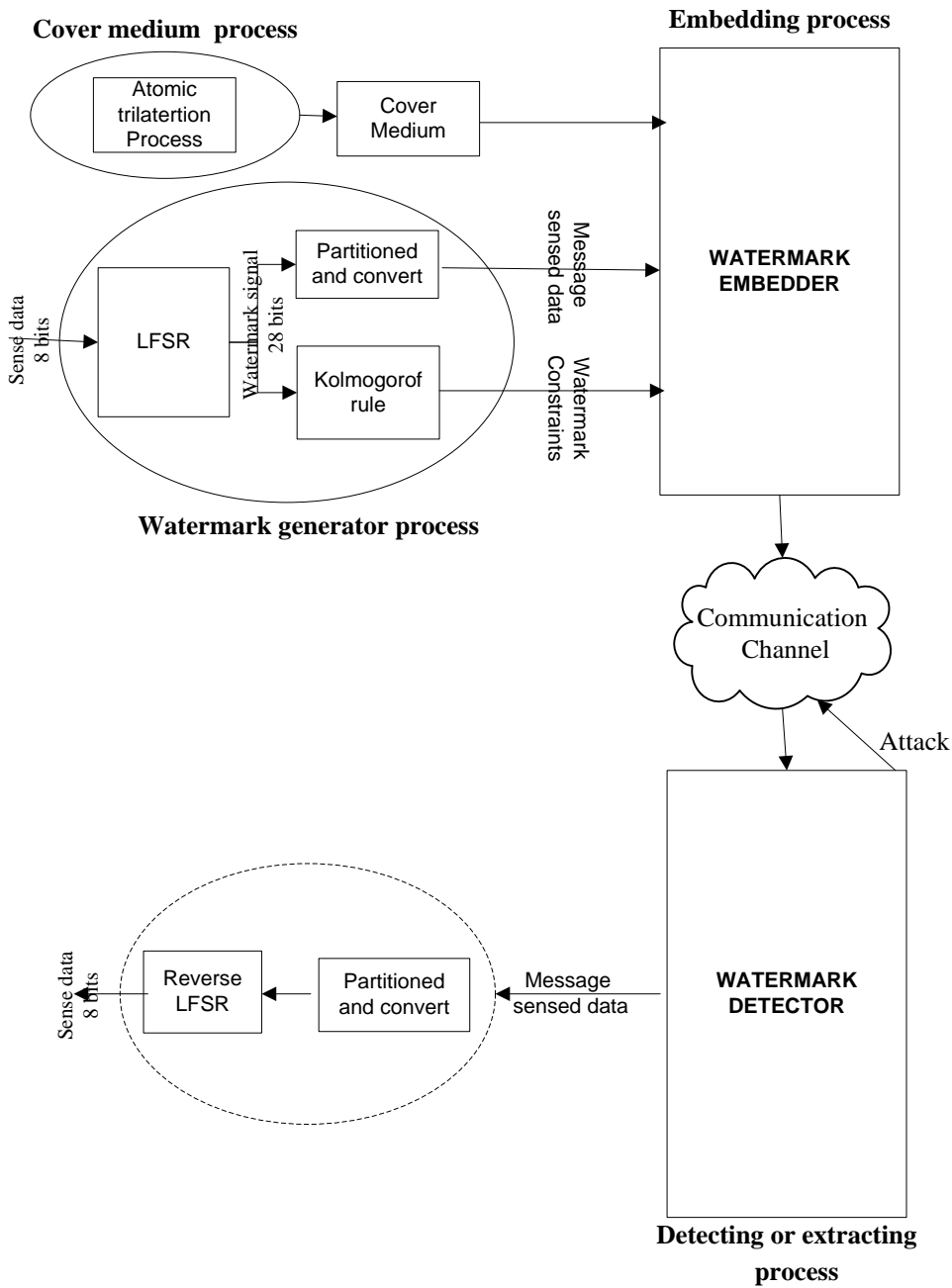

**Figure 1 secure data transmission model based on watermarking**

We next explain the four steps, we begin cover medium process

### 5.1. Generate Cover medium

In this section, we explain the process of generating cover medium by using atomic trilateration process (Pseudocode 1) With respect to a two-dimensional sensor networks, atomic trilateration is a well-known procedure by which a sensor node in a networks can determine its position by using the position of and distances to at least three other sensor nodes of know location. From these distance and position, a sensor node which is trying to determine its location can generate a nonlinear system programming.

```
Pseudocode  1. Generate Cover Medium
Input   : (x_A,y_A),(x_B,y_B), (x_C,y_C), T_c, t_DA, t_DB, t_DC,
          (x_D,y_D), ε_t, ε_DA, ε_DB, ε_DC, δ_1, δ_2, δ_3
Output: The cover medium is
```

$$\min f = \varepsilon_t + \varepsilon_{DA} + \varepsilon_{DB} + \varepsilon_{DC} + \delta_1 + \delta_2 + \delta_3$$

Constraints

$$(\sqrt{(x_D - x_A)^2 + (y_D - y_A)^2 + (z_D - z_A)}) - (331.4 + 0.6(T_c + \varepsilon_t))(t_{DA} + \varepsilon_{DA})) \leq \delta_1$$

$$(\sqrt{(x_D - x_B)^2 + (y_D - y_B)^2 - (z_D - z_B)}) - (331.4 + 0.6(T_c + \varepsilon_t))(t_{DB} + \varepsilon_{DB})) \leq \delta_2 \quad \text{Eq (1)}$$

$$(\sqrt{(x_D - x_C)^2 + (y_D - y_C)^2 - (z_D - z_C)}) - (331.4 + 0.6(T_c + \varepsilon_t))(t_{DC} + \varepsilon_{DC})) \leq \delta_3$$

```
Steps :
1. Compute  V_s = 331.4 + 0.6T_c
2. Compute  d_DA=V_s*t_DA, d_DB=V_s*t_DB, d_DC=V_s*t_DC.   Where d_DA d_DB  and  d_DC is
   between node D and the sensor nodes are then measured using
   TDoA.
3. Append ε_t error of measurement time to step (2)
4. Append ε_DA, ε_DB, and ε_DC errors of measurement distance to step
   (2).
5. Compute  d_DA = √((x_D-x_A)² + (y_D-y_A)²)  d_DB = √((x_D-x_B)² + (y_D-y_B)²)
   d_DA = √((x_D-x_C)² + (y_D-y_C)²)
6. Append δ_1, δ_2 and  δ_3  errors between the Euclidean distances step
   (3)
7. Replacing d_DA, d_DB and  d_DC from step (2) to step (3) and then
   compute them.
8. Print cover medium
```

### 5.2. Watermark generation process

Generation process is the first step and a very critical of the process. The requirements of watermark generation process are unique and complexity. The watermark message contains information that must be unique such as text and sensed data. The watermark key is also unique in order to make a secrecy key such as binary stream, integer and amplitude. Both the watermark message and the watermark key generator are as input and they then are processed in the watermark generator to produce a watermark signal. The process of generate watermark signal consists of five steps: (1) converting sensitive data into binary sequence, (2) Linear Feedback Shift Register ( LFSR ) to create watermark signal, (3) Kolmogorof rule to produce watermark constraints, (4) Partitioning and convert to decimal number from watermark signal to produce message sensed data and. Let we explain each of steps

### 5.2.1. Converting Sensitive data into binary sequence

The first step is that converting sensitive data into binary sequence. Any data of which the compromise with respect to confidentiality, integrity, and/or availability could have a material adverse effect on coventry interest, the conduct of agency programs. This data is called a sensitive data. The sensitive data is directly proportional to the materiality of a compromise of the data with respect to these criteria. Shih, F et.al [25] present finding sensitive data and privacy issue of applications in Body Sensor Networks(BSN). In BSN, the applications collect sensitive physiological data of the user and send to other parties for further analyses. The sensitive data are heart rate and Blood Pressure. These data are required to be protected and then these data will be converted scalar data into binary stream. WSN has gathered a blood pressure patient. The patient blood pressure is 120 so the digit sequence of 120 padded with zeros so that it is of total length 8. d= dec2bin([120],8) = 01111000

### 5.2.2. Generating watermark signal using LFSR

One method of forming a binary sequence for generating watermark is to apply a LFSR whose characteristic polynomial is primitive [26, 27]. LFSR is a shift register whose input bit is a linear function of its previous state. The only linear function of single bits is exclusive-or (*xor*), therefore it is a shift register whose input bit is driven by *xor* of some bits of the overall shift register value.

LFSR can be defined by a recurrence relation:

$$s_{K+n} = \sum_{i=0}^{n-1} c_i s_{k+1}, \text{ where } k \geq 0, n \in Z \text{ and the } c_i \text{ are binary constants such that } c_O = 1.$$ , Eq (2)

associated with such a recurrence relation is a binary polynomial

$$f(x) = c_0 + c_1 x + ... + c_{k-1} x^{k-1} + x^k,$$ Eq (3)

called the characteristic polynomial of the LFSR. The coefficient $c_i$ are feedback constants. Such sequence can be mechanized by using a LFSR whose tap setting are defined by the feedback constants.

We implemented (pseudocode 2) to generate a watermark signal, we use the sensory data as the initial state of LFSR, i.e., "01111000" and the binary polynomial $f(x) = 1 + x + x^5 + x^6 + x^7$. This binary polynomial is written by [ 1 2 5 6 ] as key embedding. We then get the 28 binary sequence is 00011110 000011011100 1100 0111 .This binary sequence is called a watermark signal.

```
Pseudocode 2. Generate watermark  signal
Input  : Sensed data, coefficients c_i of the binary  polynomial
         as watermark key
Output :  28 bits watermark signal
Steps  :
1. Convert sensed data into binary sequence.
2. Use the coefficients c_i of the binary polynomial f(x) as
   watermark key
3. Generate an infinite binary sequence using the coefficient c_i
   into a LFSR (s_{K+n}).
4. The infinite binary sequence cut from 1 to 28 as watermark
   signal.
5. Print 28 bits watermark signal
```

### 5.2.3. Kolmogorov rule to create watermark constraints

Andrew nikolaevich Kolmogorov [28] states that complexity of an object is the length of shortest computer program that can reproduce the object. The Kolmogorov complexity is defined a probability distribution under which worst-case and average-case running time are the same. We know that kolmogorof rule is the short description length of overall description interpreted by computer. The three papers [5, 29, 30] used the kolmogorov rule for numbering the variables of linear combination in the optimization objective function and a set of constrains. We also use the kolmogorov rule. This rule can be seen in Table 2

**Table 2 the kolmogorov rule**

| 1 | 2 | 3 | 4 | 5 | 6 | 7 |
|---|---|---|---|---|---|---|
| $\varepsilon_t$ | $\varepsilon_{DA}$ | $\varepsilon_{DB}$ | $\varepsilon_{DC}$ | $\delta_1$ | $\delta_2$ | $\delta_3$ |

We number ( pseudocode 3) these variables by using kolmogorov rule.

```
Pseudocode3. Generate watermark  constraints
Input : 28 bits watermark signal
Output: Watermark constraints
Steps
1. Group  28 bits watermark signal into group of 7 bits each.
2. Match the bit number with corresponding variable number from
   table 2.
3. If a bit one is assigned a variable with in a group that
   variable is included in the linear
4. Else a bit zero is assigned a variable with in a group that
   variable is not included in the linear.
5. Go to 2
6. Print watermark constraints
```

### 5.2.4. Partitioned and convert to create message sensed data

In this section, we explain how a message sensed data created. To create this message sensed data ( pseudocode 4), 28 bits watermark binary that resulting from generating watermark signal is used.

```
Pseudocode 4. Generate create message data
Input : 28 bits watermark signal
Output: message sensed data
Steps:
1. Group 28 bits watermark signal into group of 4 bits each.
2. Convert each of group into decimal number to get weight
   factors.
3. Print message sensed data
```

### 5.3.    Watermark Embedding process

Embedding process is the second step of the watermarking system that is undertaken by an watermark embedder. The embedder combines the cover medium, the watermark constraints and the message sensed data and it then creates watermarked cover medium. The watermarked cover medium is perceptually identical to the cover medium. The figure 2 shows the watermark embedding process. The watermark signal is converted to become watermark constraints by using kolmogorov rule. The watermark constraints consist of four constraints that will be added into the Equation 1. The message sensed data is also inserted into the coefficient objective of Equation 1. This message sensed data is a weight factors that obtained by partitioning the watermark signal into 7 sections. The watermark signal is converted to become watermark constraints by using kolmogorov rule. The watermark constraints consist of four constraints that will be added into the Equation 1. This message sensed data is a weight factors that obtained by partitioning the watermark signal into 7 sections. The redundant constraints are added into the Equation 1. The watermarking embedding process ( Pseudocode 5) can be shown in Figure 2.

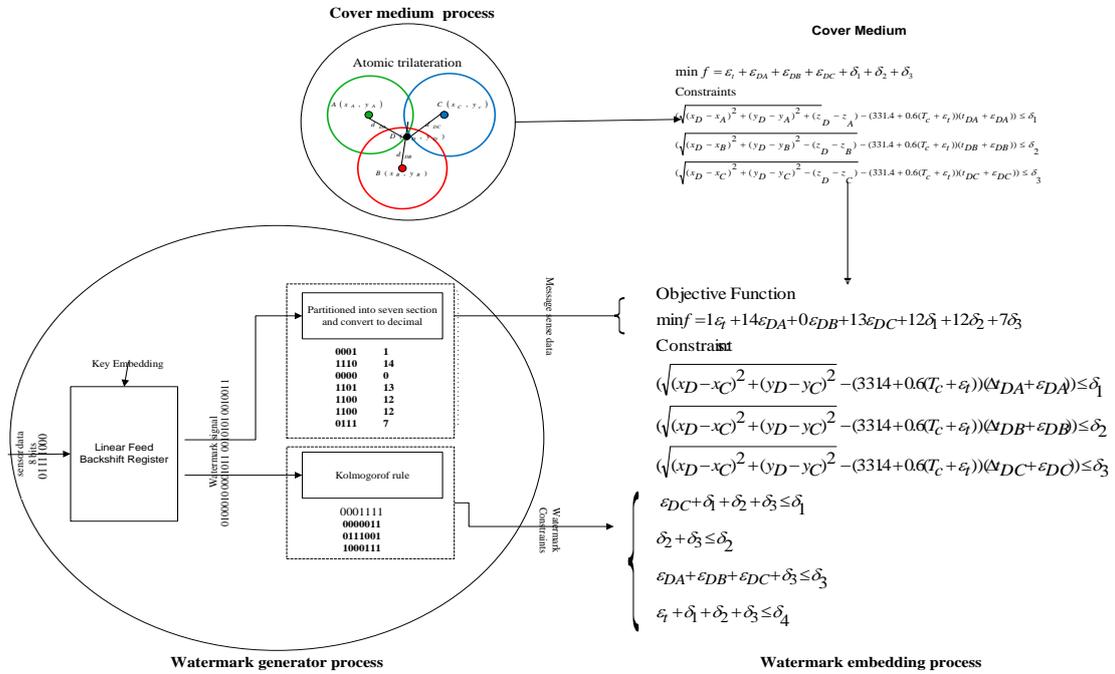

**Figure 2 Watermark Embedding Process**

```
Pseudocode 5. The process of embedding
Input:  cover medium, Watermark constraints, message sensed
 data.
Output: (x_D, y_D), ε_t, ε_DA, ε_DB, ε_DC δ_1, δ_2, δ_3
      and min f
Steps :
1. Generate (x_A,y_A),(x_B,y_B) and (x_C,y_c) using uniform distribution on
   interval [0,1].
2. Generate t_DA, t_DB, and  t_DC using uniform distribution on interval
   [0.02,0.1].
3. Generate δ_1,δ_2 and  δ_3 using gauss distribution on interval
   [0,1].
4. Generate τ_1,τ2,τ_3 and τ_4 using gauss distribution on interval
   [0,1], So that these value do not harm to the feasibility of
   the solution of the  cover medium
5. Generate T_c using gauss distribution on interval  [0,1].
6. Change coefficient objective f to weight factor of message
   sensed data respectively.
7. Append watermark constraints into cover medium
8. Compute and print (x_D,y_D), ε_t, ε_DA, ε_DB, ε_DC δ_1,δ_2, δ_3  and min f
```

## 5.4. Watermark detecting and extracting process

The process of detecting watermark into has not yet explained in Feng Jasica P et.al [5] and F. Koushanfar et.al [8]. Both of them are only explain the process of embedding watermark. To verify the presence of the watermark, we adopt the concept of Cox *et al* [31]. Cox draw parallels between their technology and spread-spectrum communication since the watermark is spread over a set of visually important frequency components Let *X* be the error from the optimal solution without watermark and *X'* be the error form the optimal solution with watermark. For detecting the watermark, a correlation value or similarity measure is used in most of these methods. Here to verify the presence of the watermark constraints, the similarity measure between the normalized difference error from the optimal solution between the watermarked solution and the solution obtained without watermarked $C' = X' - X$. Adding the message sensed data into the Equation 1 is called the equation without watermark constraints. Adding the message sensed data and the watermark constraints are called the Equation 1 with watermark. The similarity measure is given by the normalized correlation coefficient $sim(C', X') = \frac{C'.X'}{\sqrt{X'.X'}}$. Subsequently, since the expected result is dyadic (i.e. the cover medium '*is*' or '*is not*' watermark), some kind of threshold is needed. The watermarking detecting process can be shown in Figure 3. This process (**Pseudocode 6**) is also can be used to obtain the value of threshold. This threshold is extracted by statistical rules and usually has a strong mathematical formulation. There are two kinds of errors in such schemes. *False-positive* corresponds to the case of detection of non-existing watermarks signal. *False-mark* stands for the case that the watermark signal exists but cannot be detected. Although well-reasoned, the existing thresholds many times lead to false-negative errors. We use False-negative to determine the watermark signal presents or not.

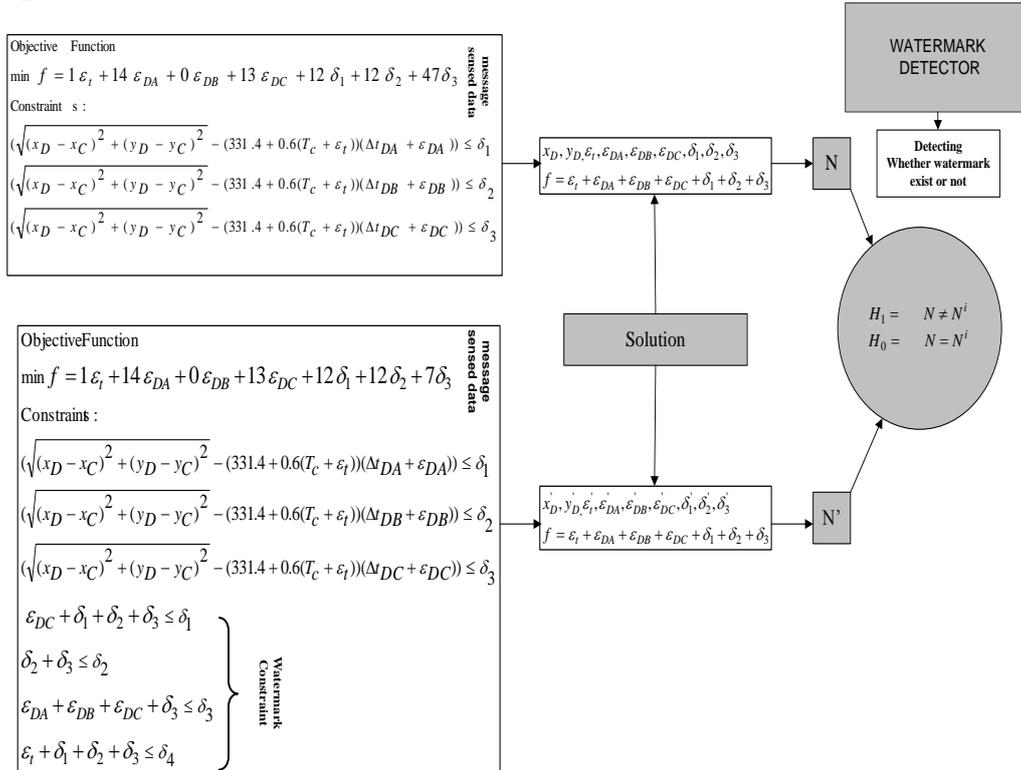

**Figure 3 Watermark detecting process**

$H_1 = \quad N \neq N'$ the cover medium is watermarked

$H_0 = N = N'$ the cover medium is not watermarked

**Pseudocode 6. The process of detecting**

**Input** : x=[ $\varepsilon_t, \varepsilon_{DA}, \varepsilon_{DB}, \varepsilon_{DC}, \delta_1, \delta_2, \delta_3$ ], x'=[ $\varepsilon_t', \varepsilon_{DA}', \varepsilon_{DB}', \varepsilon_{DC}', \delta_1', \delta_2', \delta_3'$ ] and x"=[ $\varepsilon_t'', \varepsilon_{DA}'', \varepsilon_{DB}'', \varepsilon_{DC}'', \delta_1'', \delta_2'', \delta_3''$ ]

**Output** : Watermark signal robust or not robust

Steps :
1. Compute $N = |[\varepsilon_t, \varepsilon_{DA}, \varepsilon_{DB}, \varepsilon_{DC}, \delta_1, \delta_2, \delta_3]|$ , $N' = |[\varepsilon_t', \varepsilon_{DA}', \varepsilon_{DB}', \varepsilon_{DC}', \delta_1', \delta_2', \delta_3']|$ and $N'' = |[\varepsilon_t'', \varepsilon_{DA}'', \varepsilon_{DB}'', \varepsilon_{DC}'', \delta_1'', \delta_2'', \delta_3'']|$
2. Compute c=x'-x and c'=x"-x
3. Compute normalized correlation the results of error the cover medium without watermark constraints $treshold = \frac{C.X'}{\sqrt{X'.X'}}$ )
4. Compute normalized correlation the results of error the cover medium with watermark constraints $sim(C', X') = \frac{C'.X'}{\sqrt{X'.X'}}$ .
5. If $N \neq N^i$ watermark signal exits
6. If $N = N^i$ watermark signal does not exist
7. If threshold $treshold < sim(C', X')$ watermark signal is robust go to 9
8. If threshold $treshold < sim(C', X')$ watermark signal is not robust
9. Algorithm the process of extracting message sensed data

The extracting process (Pseudocode 7) is also undertaken in the watermark detector, we want to recovery the message sensed data from the cover medium. Based on the statistical rule of false-negative, we accept $H_1$ that means the cover medium is watermarked.

We then can do the process of extracting a watermark message sensed data into sensed data as shown in Figure 4. By using the pseudo code 7 the value of errors the cover medium with watermark constraints, we check whether these watermark constraints do not change or not. If these constraints do not change, we can do the process of extracting watermark signal. In this case, the coefficients objective function form the cover medium are 1 14 0 13 12 12 and 7.

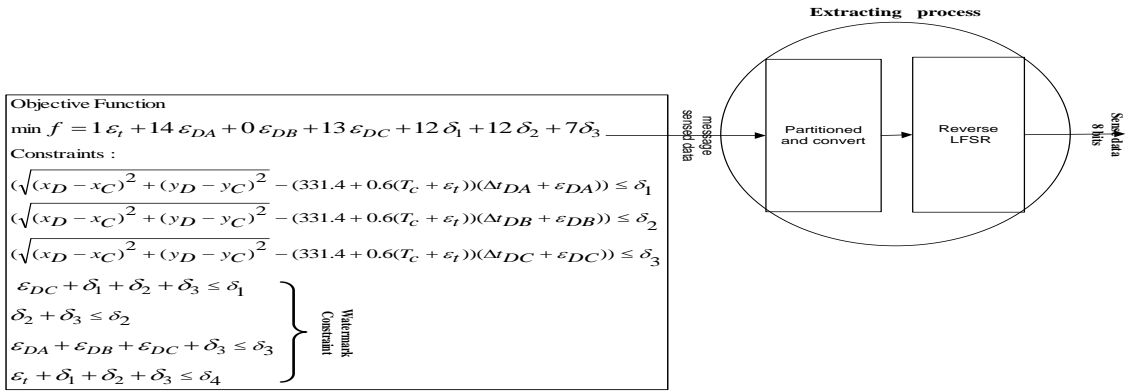

**Figure 4 Watermark extracting Process**

**Pseudocode 7. The process of extracting sensed data**

Input : $\varepsilon_t, \varepsilon_{DA}, \varepsilon_{DB}, \varepsilon_{DC}\ \delta_1, \delta_2, \delta_3$ watermark key
Output : Sensed data
**Steps :**

1. Compute the value of the objective f using $\varepsilon_t, \varepsilon_{DA}, \varepsilon_{DB}, \varepsilon_{DC}\ \delta_1, \delta_2,$ and $\delta_3$
2. If the value of the objective do not change go to 3
3. Else the value of the objective change goes to step 1.
4. Take the coefficients of objective f.
5. Convert the coefficient of objective f into 4 bits each.
6. Merge all of these 4 bits to 28 bits
7. Use reverse LFSR with watermark key to get sensed data.

## 6. EXPERIMENT SETUP

In this section, we describe the experiment setup for testing the purpose of the secure data transmission model, based on watermarking technique. We used TOMLAB which is a general purpose development environment in MATLAB for research and practical solution of optimization problems. TOMLAB has grown out of the need for advanced, robust and reliable tools to be used in the development of algorithms and software for the solution of many different types of applied optimization problems

### 6.1. NETWORK SETUP

In this section, the scenario of the atomic trilateration process is used as shown in Figure 4

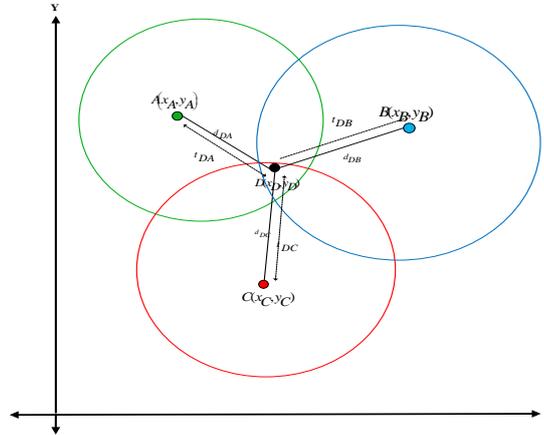

**Figure 4 Atomic trilateration**

With respect to a two-dimensional sensor networks, atomic trilateration is the means by which a sensor node in a networks can be used to determine its position by using the position of and distances to at least three other multimedia sensor nodes of know location. From these distance and position, a multimedia sensor node which is trying to determine its location can generate a nonlinear system equation. A typical scenario of atomic trilateration can be shown in Figure 4. Sensor node *D* trilaterates with another three sensor nodes *A*, *B*, and *C* which have coordinates $(x_A, y_A), (x_B, y_B),$ and $(x_C, y_C)$. The distance is computed using time differences of arrival (TDoA) between acoustic signals simultaneously, which are emitted from a sensor nodes and received at the node *D* and radio frequency (RF). The sensor node *D* turns on a timer upon receiving the RF signal from the sensor node to measure the difference between the arrival of the RF and acoustic signals from that sensor node. The time measurements have an error. The speed of the acoustic signal is a function of the temperature of the propagation media. The relationship between the speed of the acoustic signal *Vs* (m/s) and the temperature $T_c$ is as follows:

$$V_s = 331.4 + 0.6 T_c \quad \text{Eq.(4)}$$

By using the pseudocode 1, we find that the objective function is to minimize the overall error in the system, and can be stated as shown in Equation (1)

## 6.2. PERFORMANCE METRICS

The existing performance of the watermarking technique for secure data transmitting is evaluated against the following performance metrics:

**Table 3 Performance Metrics secure data transmitting**

| Parameter | explain | Metric | Value |
|---|---|---|---|
| Node Sensor | Number of sensor node | Integer | 100 |
| $(x_i, y_j)$, $i = j = 1,2,...n$ | Position of two-dimensional sensor networks | Coordinate | $x_i = 115,5693$ $y_i = 273,2856$ |
| $T_c$ | the temperature of the propagation media | Degree | $T_c = 36$ |
| $t_{DA}, t_{DB}, t_{DC}$ | time transmission between node D to A, D to B and D to C | second | $t_{DA} = 0,771625$ $t_{DB} = 0,106793$ $t_{DC} = 0,09282$ |
| *Vs* | Speed acoustic signal | (m/s) | $V_s \geq 331.4$ |
| $\varepsilon_t$ | the error in the measurement of the temperature | - | $\varepsilon_t = 0$ |
| $\varepsilon_{DA}, \varepsilon_{DB}, \varepsilon_{DC}$ | the error in the measurement of the timer from D to A, D to B and D to C | - | $\varepsilon_{DA} = 0.0473$ $\varepsilon_{DB} = -0.0141$, $\varepsilon_{DC} = 0$ |
| $\delta_1, \delta_2, \delta_3$ | the error in the measurement between the Euclidean measurement and the measured using time differences of optimal D to A, D to B and D to C. | - | $\delta_1 = 0, \delta_2 = 0, \delta_3 = 0$ |
| $\tau_1, \tau_2, \tau_3, \tau_4$ | the values are selected such that the feasibility of the solution space of the optimization problem is not harmed | - | $\tau_1 = 0.16947616$ $\tau_2 = 0.16947616$, $\tau_3 = 0.24915965$ $\tau_4 = 0.992920660$ |
| Sensed data | Data sensed by a sensor node | Bit | 01111000 |
| Watermark signal | Result from LFSR | Bit | 00011110 000011011100 1100 |
| Message sensed data | Result from pseudo code 4 | Integer | 1 14 0 13 12 12 7 |
| *treshold* | normalized correlation the results of error the cover medium with watermark constraints | - | 0.799153536405721 |
| $sim(C', X')$ | normalized correlation the results of error the cover medium with watermark constraints attack | - | 0.2.154207742903002 |

# 7. EXPERIMENT AND RESULTS FOR SECURE DATA TRANSMITTING

WSNs have an additionally vulnerability because node are often placed in a hostile environment where they are not physically protected. An attack is considered successful if it is not detected by the receiver. In this section we discuss various types of attacks that can be launched in the WSNs scenario and how the proposed security scheme can be used to thwart these attacks.

We consider in detail the corresponding weakness for this model watermarking technique that could be used by the attacker. Assume that the watermarks constraints are estimated by the attacker that should be change, modify and remove. The corresponding attacks are:

## 7.1 False data insertion attack

A number of different watermarks constraints that are generated by the LFSR, hoping to find the new results of error the cover medium that will map into existing solution.

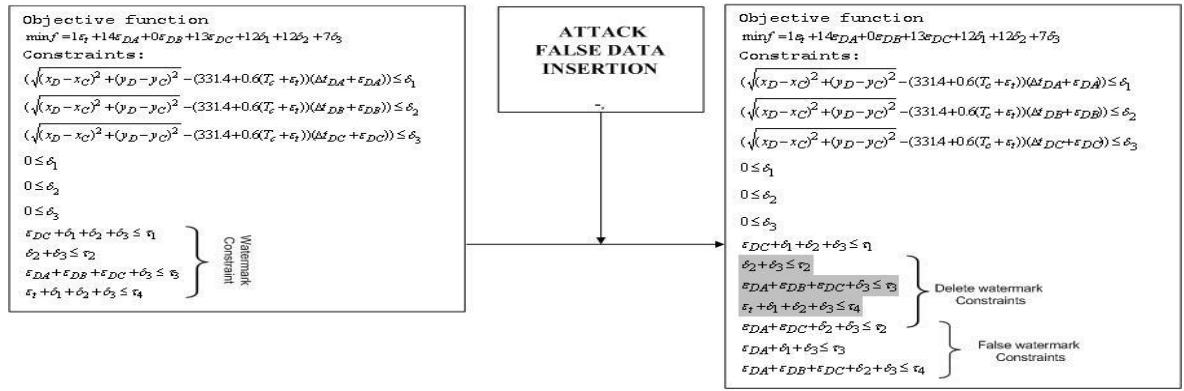

We get the results of the error of the cover medium by false insertion watermark constraints: $\varepsilon_t = 2.154233085444387$, $\varepsilon_{DA} = 0.007135532211399$, $\varepsilon_{DB} = -0.000927368803587$, $\varepsilon_{DC} = 0.001724459319967$, $\delta_1 = \delta_2 = 0$ and $\delta_3 = 0$.

Implementing a pseudo-code 6, we conclude that the value of similarity is greater than the value of threshold: the value of similarity = $2.154207742903002e+002$ > the value of threshold = $0.799153536405721$. This means that the watermark signal is not robust to false data insertion attack.

## 7.2 Data modification attack

Data modification attack makes impersonation of different watermarks constraints that are generated by the LFSR, hoping also to find the new results of error the cover medium that will map into existing solution.

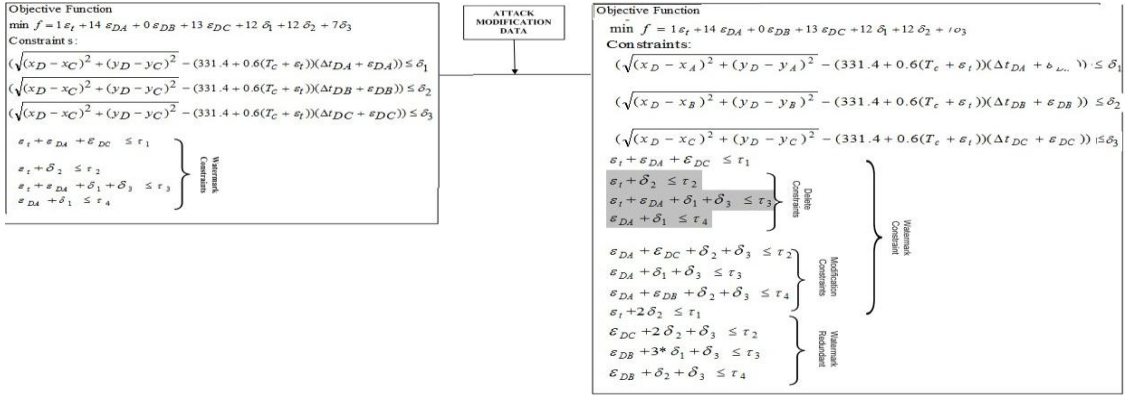

We get the results of the error of the cover medium by modification watermark constraints: $\varepsilon_t = 0.100170911928198$, $\varepsilon_{DA} = 1.118559233568045$, $\varepsilon_{DB} = -0.167216683069220$, $\varepsilon_{DC} = 0.000000000000137$, $\delta_1 = \delta_2 = 0$ and $\delta_3 = 0.145088180096586$

Implementing a pseudo-code 6, we conclude that the value of similarity is greater than the value of threshold: the value of similarity $= 0.923139703988680 >$ the value of threshold $= 0.799153536405721$. This means that the watermark signal is not robust enough to modification the attack.

### 7.3 Data Deletion Attack

Data deletion attack is similar to the spoofed data attack in the sense that deleting watermark constraints make the error results of the cover medium invalid Delete a number of watermarks constraints hope to find new results of error the cover medium.

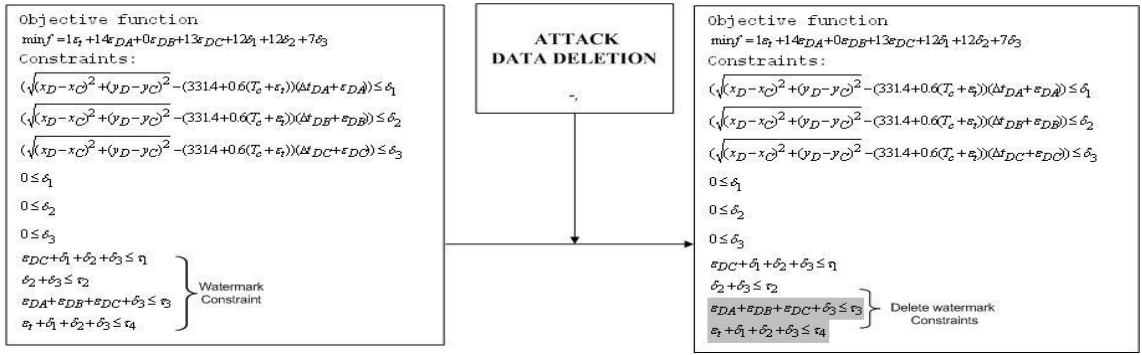

We get the results of the error of the cover medium by deleting watermark constraints: $\varepsilon_t = 0$, $\varepsilon_{DA} = 0.931857673282008$, $\varepsilon_{DB} = -0.870667531648967$, $\varepsilon_{DC} = 0.145088180096586$, $\delta_1 = \delta_2 = 0$ and $\delta_3 = 0$.

Implementing a pseudo-code 7, we conclude that the value of similarity is greater than the value of threshold: the value of similarity $= 0.352844500181367 <$ the value of threshold $= 0.799153536405721$. This means that the watermark signal is robust enough to delete the attack.

### 7.4 Replication Attack

Data replication attack is quite simple: an attacker seeks to add new constraints to the cover medium by replicating the new constraints with the existing constraints. New constraints replicated in this fashion can severely disrupt this solution of the cover medium's performance. Data replication attack hopes to find the new results of error the cover medium that will map into existing solution.

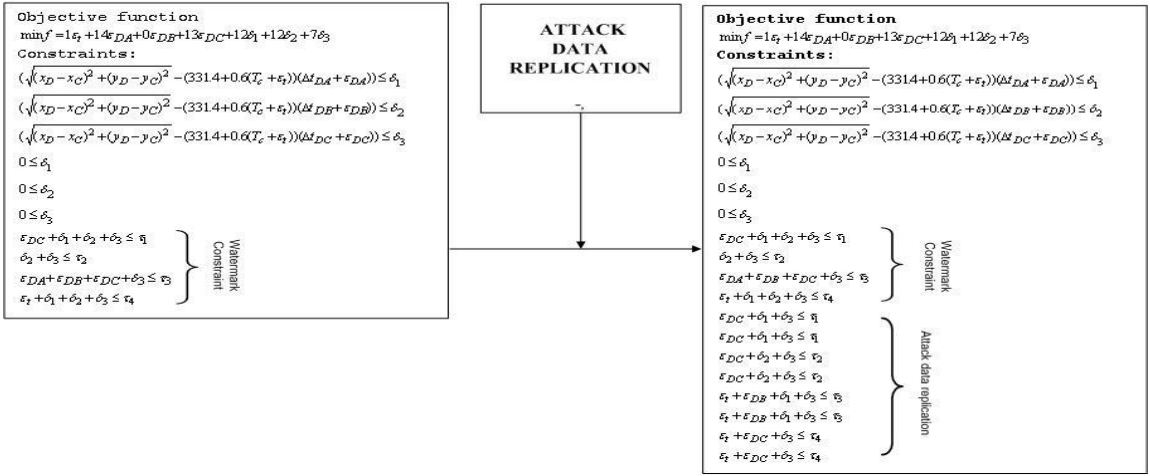

We get the results of the error of the cover medium by replication watermark constraints: $\varepsilon_t =0.122299414900832$, $\varepsilon_{DA} = 0.428473573865247$, $\varepsilon_{DB} = -0.500740473660944$, $\varepsilon_{DC} = 0$, $\delta_1 = \delta_2 = 0$ and $\delta_3 = 0.145088180096586$.

Implementing a pseudo-code 7, we conclude that the value of similarity is greater than the value of threshold: the value of similarity = 0.285586856097203 < the value of threshold = 0.799153536405721. This means that the watermark signal is robust to Replication Attack.

### 7.5 Sybil attack

A Sybil attack data occurs when the attacker creates multiple identities and exploits them in order to manipulate a reputation score. The Sybil attack data is defined as a malicious device illegitimately taking on multiple data identities., The Sybil attack data in communication channel watermarking is an attack wherein a reputation network system is subverted by forging more than one identity constraints in the cover medium A Sybil hopes to find aresults of error the cover medium.

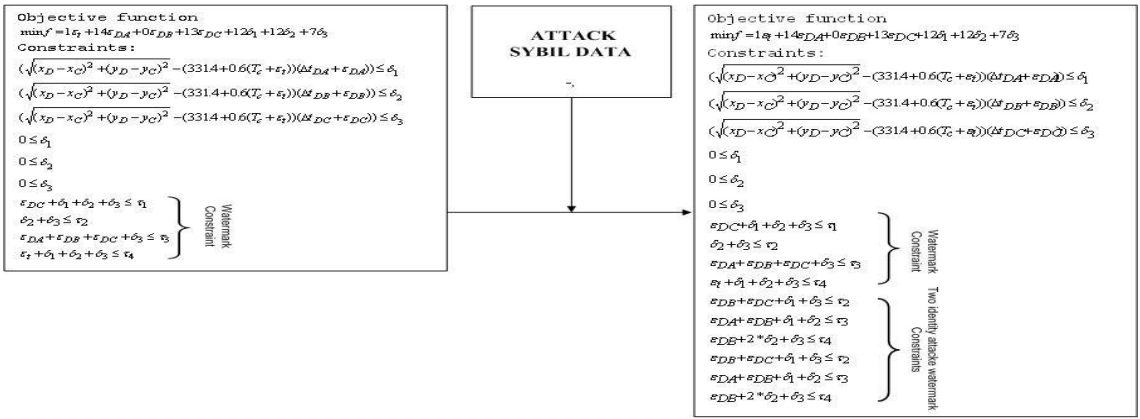

We get the results of the error of the cover medium by Sybil watermark constraints: $\varepsilon_t = 0.100170911928198$, $\varepsilon_{DA} = 0.118559233568045$, $\varepsilon_{DB} = -0.013888456874134$, $\varepsilon_{DC} = 0.000000000000137$, $\delta_1 = \delta_2 = 0$ and $\delta_3 = 0$.

Implementing a pseudo-code 7, we conclude that the value of similarity is greater than the value of threshold: the value of similarity = 0.103640805769825 < the value of threshold = 0.799153536405721. This means that the watermark signal is robust enough to Sybil attack the attack.

The results of these experiments have been shown in Table 3.

Table 3  The robustness of a watermark constraints, and watermark signal

| No. | Kind of attacks | Watermark constraints | Watermark Signal |
|---|---|---|---|
| 1. | False data insertion | Change | Not robust |
| 2. | Data modification attack | Change | Not robust |
| 3. | Data deletion | Not change | Robust |
| 4. | Packet replication. | Not change | Robust |
| 5. | Sybil attack | Not change | Robust |

## 8. PERFORMANCE EVALUATION

In this section, we perform a comparative analysis of our technique with other techniques proposed by different researchers. The results of this comparative analysis are given in Table 4

Table 4  Comparative analysis with other approach

| Kind of attacks | Feng et al. [5] | Sion et al. [6] | Koushanfar et al. [8] | Zhang et al. [10] | Xiao et al. [9] | Xuejun et al. [11] | Kamel et al. [13] | Harjito B |
|---|---|---|---|---|---|---|---|---|
| False data insertion | x | x | x | √ | x | x | √ | x |
| data modification attack | x | x | x | √ | x | x | √ | x |
| Data deletion | x | x | x | x | x | x | √ | √ |
| Packet replication. | x | x | x | x | x | x | x | √ |
| Sybil attack | x | x | x | x | x | x | x | √ |

√ provide secure data communication and robust    x not provide secure data communication

We then do many experiments of these attacks above to test the performance of the model of secure data communication in WSNs. The results of these experiments can be shown in Table 3.

In this works, we compare 8 approaches in term of false data insertion, data modification attack, data deletion, packet replication, and Sybil attack. The [5], [6], [8], [9] approaches do not provide secure data communication against 6 attack. But [10] provide data for copy right protection and [13] provide for data integrity against these attacks.  Our approach provides secure data communication against data deletion, packet replication and Sybil attacks. However our approach does not provide secure communication against false data insertion, and modification data

## 9. CONCLUSIONS

In this paper, we propose a watermarking technique for secure data transmitting in WSNs. Our strategy aims to protect data transmitting between sensor nodes in WSNs against these attacks. We verify our technique by brute force attacks. We can make secure data from data deletion, packet replication and Sybil attacks. However we cannot protect secure data from false data insertion, and modification data. Therefore, we still need to improve our technique under the circumstance that attacker launch different attack for the future work.

**Authors**

**Bambang Harjito** is now as head of computer science department at Mathematics and Natural Science, Sebelas Maret University Surakarta, Indonesia. He received the master degree in computer science department at James Cook University in 2000 and He received PhD in School of information System, Curtin University Perth Australia in 2013.

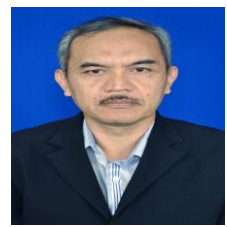

**Vidyasagar Potdar is** a Senior Research Fellow working with School of Information Systems, Curtin Business School, Curtin University, Perth, Western Australia. He received the Bachelor of Science, Gujrat University, India 2001 and the Master of Informatoin Technology - University of Newcastle, Australia in 2002 and Doctor of Philosophy - Curtin University of Technology, Australia 2006. He is the Director of Anti-Spam Research Lab & Co-Director of Wireless Sensor Network Lab at the School of Information Systems.

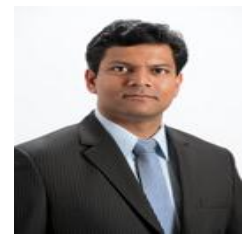